\documentclass[10pt, conference]{FMFP2024}
\usepackage[top=1.50cm, bottom=1.50cm, left=1.884cm, right=1.884cm, includehead, includefoot, heightrounded]{geometry}
\usepackage{fancyhdr} 
\usepackage{graphicx} 
\usepackage{epstopdf,epsfig}
\usepackage[textfont=bf, font = normalsize]{caption} 
\usepackage{floatrow} 
\usepackage{amssymb}
\usepackage{amsfonts}
\usepackage{amsmath}
\floatstyle{plaintop} 
\restylefloat{table} 
\usepackage{subfigure} 
\renewcommand{\headrulewidth}{0pt}

\begin{document}

\title{\LARGE{ \bf Detection and analysis of synchronization routes in an axially forced globally unstable jet using recurrence quantification}}

\author{\textbf{Abhijit K. Kushwaha\textsuperscript{1,4,*}, 
Meenatchidevi Murugesan\textsuperscript{2}, Nicholas A. Worth\textsuperscript{3}, James R. Dawson\textsuperscript{3}},\\
\textbf{Tadd T. Truscott\textsuperscript{1}, and Larry K. B. Li\textsuperscript{4}}\\
\textsuperscript{1}\small{Department of Physical Science and Engineering, King Abdullah University of Science and Technology, Thuwal, Saudi Arabia}\\
\textsuperscript{2} \small{Department of Mechanical, Materials and Aerospace Engineering, Indian Institute of Technology, Dharwad, India}\\
\textsuperscript{3} \small{Department of Energy and Process Engineering, Norwegian University of Science and Technology, Trondheim, Norway}\\
\textsuperscript{4} \small{Department of Mechanical and Aerospace Engineering, The Hong Kong University of Science and Technology, Hong Kong}\\
\small{* Corresponding Author, email: abhijitkumar.kushwaha@kaust.edu.sa}}

\maketitle
\thispagestyle{fancy} 
\pagestyle{plain} 

\noindent \textbf{ABSTRACT}

Quasiperiodicity, a partially synchronous state that precedes the onset of forced synchronization in hydrodynamic systems, exhibits distinct geometrical patterns based on the specific route to lock-in. In this study, we explore these dynamic behaviors using recurrence quantification analysis. Focusing on a self-excited hydrodynamic system—a low-density jet subjected to external acoustic forcing at varying frequencies and amplitudes—we generate recurrence plots from unsteady velocity time traces. These recurrence plots provide insight into the synchronization dynamics and pathways of the jet under forced conditions. Further, we show that recurrence quantities are helpful to detect and distinguish between different routes to lock-in.


\section{{\textbf{INTRODUCTION}}}
Open jet flows are encountered in both natural and engineering systems. A globally stable jet acting as spatial amplifier of external disturbances can become globally unstable if the jet contains a sufficiently large region of local absolute instability \cite{Chom88,Monk90}. A globally unstable jet exhibits self-excited oscillations with an intrinsic natural frequency, $f_n$ \cite{Hall06,lesshafft2007}. A self-excited jet adapt its rhythm to synchronize with the forcing signal when the jet is forced at a frequency $f_f$ different from its natural frequency $f_n$ \cite{balanov2008}. Depending on the the forcing amplitudes, $A$ and freqeuncy, $f_f$, the forced jet exhibits a wide range of nonlinear dynamics and bifurcations. Specifically, the jet exhibits quasiperiodic oscillations when forced below a critical forcing amplitude \cite{pikovsky2003}. However, when it is forced above a critical amplitude, the jet lock-in to the forcing signal \cite{staubli1987JFM}. 

The onset of lock-in gives rise to a forced synchronous state. Forced synchronization has been extensively studied as a promising open-loop flow control strategy in various hydrodynamically self-excited systems such as bluff-body wakes \cite{schumm1994}, capillary jets \cite{Olin92}, axisymmetric low-density jets \cite{kushwaha2022}, cross-flow jets \cite{Davi10b}, diffusion flames \cite{rick2019open}, and thermoacoustic systems \cite{mondal19}. To develop an improved flow control strategies relies on the ability to characterize, understand, and predict the forced synchronization characteristics and bifurcations leading up to lock-in of hydrodynamically self-excited flows. 

In the framework of forced synchronization, a self-excited system forced at an off-resonance frequency can lock-in to $f_f$ via a two universal routes to: (i) phase-locking and (ii) suppression \cite{balanov2008}. A self-excited system locks into $f_f$ via a phase-locking route when the detuning frequency ($|f_f-f_n|$) is small, which occurs when $f_f$ is close to $f_n$. By contrast, when $f_f$ is far from $f_n$ for which the detuning frequency is large, the lock-in occurs via suppression route. A key attribute of these two different lock-in routes are their unique bifurcation leading upto lock-in. Along the phase-locking route, the system undergoes saddle-node bifurcation during which $f_n$ is gradually pulled towards $f_f$ without ever reduction in the amplitude of $f_n$ mode as the forcing amplitude increases. The suppression route to lock-in, however, occurs when the system exhibits an
inverse Neimark–Sacker bifurcation, also known as tours-death bifurcation during which the $f_n$ mode gradually weakens without any gradual shift towards $f_f$ as the forcing amplitude increases.

Regardless of whether the forced self-excited systems undergo saddle-node or torus-death bifurcation, the transition to lock-in state always occurs from a quasiperiodic state. Quasiperiodicity is a common attributes of a weakly forced self-excited system, resulting from the interaction of at least two periodic modes with incommensurate natural frequencies, $f_f$ and $f_n$ \cite{Thom02}. Consequently, a quasiperiodic system displays aperiodic dynamics with oscillations that repeat over an infinite period. This leads to an emergence of $\mathbb{T}^2$ torus attractor in the phase space via Neimark–Sacker bifurcation. Although the transition from a periodicity when unforced to a quasiperiodicity for a weakly forced system occurs via same bifurcation route, the stability characteristics of the torus attractor depends on the detuning frequency or the specific synchronization route to lock-in, which collectively determines the suppression or amplification of overall the oscillations amplitude . 

For a robust control, it is crucial to determine the how much the $f_f$ should be tuned away from $f_n$ and track the specific bifurcation in the lead up to lock-in that results in the suppression of overall oscillations amplitude. Instead of relying on conventional methods like tracking changes in time traces, frequency spectra, or phase space—which can lead to misinterpretation in route identification—we introduce new approaches to complement existing tools for diagnosing and predicting synchronization boundaries and specific synchronization routes in this paper.


\section{\textbf{Experimental setup}}\label{sec2}

The experimental set up in this study is identical to that in our previous work by Kushwaha et. al. \cite{kushwaha2022}, only a brief overview is provided here. The setup consists of a convergent nozzle with a round outlet, having an exit diameter of $D = 6$ mm and a contraction ratio of 34:1. When helium gas is discharged from the nozzle outlet, it produces a globally unstable axisymmetric low-density jet. At Reynolds number of $Re =800$ and the transverse curvature of $D/\theta_o$ is 33.2, the jet is globally unstable to axisymmetric mode, $m=0$ \cite{Jend94}. The jet exhibits self-excited oscillations at a discrete natural frequency of $f_n=945$ Hz $\pm 15 \%$ at $Re=800$. The jet is axially forced when placed at the pressure anti-node of planar acoustic standing waves, which are generated in a rectangular enclosure housing a pair of loudspeakers (Monacor KU-516) mounted at opposite ends. The jet is forced over a wide range of frequencies  ($0.84 \leq f_f/f_n \leq 1.16$) to explore the routes to forced synchronization. At each forcing frequency ($f_f/f_n$), the axial forcing amplitude $A$ (maximum pressure oscillation at the nozzle outlet centerline) is increased incrementally. At each forcing condition ($f_f/f_n$ $\&$ $A$),  we measure the jet response in terms of unsteady velocity using a hot-wire anemometer. The hot-wire anemometer consists of a probe made of a 5 $\mu$m diamtere tungsten wire and it is operated in a constant-temperature mode at an overheat ratio of 1.8. The probe is calibrated for both air and helium gas by following the procedure in \cite{John05} to a uncertainty of $\pm1.7\%$ at 95 \% confidence on the normal distribution. We measure the unsteady velocity along the jet centreline at 1.5$D$ downstream of the jet exit, at a sampling rate of 25600 Hz for a duration of 8 seconds. The schematic of the configuration and further details on experiments can be obtained from \cite{kushwaha2022}.

\section{\textbf{Time-series analysis based on recurrence quantification}}\label{sec3}

\subsection{\textbf{Recurrence plots}}

In the present study, we identify dynamical states, synchronization dynamics, and patterns in time series data of a forced self-excited jet based on the fundamental concept of recurrence of a trajectory in phase space \cite{marwan2007}. Eckmann, Kamphorst $\&$ Ruelle \cite{eckmann1995} used the recurring behavior of dynamical systems to develop a graphical tool known as recurrence plots. It is represented as a two-dimensional binary bitmap, with its elements defined by the recurrence matrix, $R_{i,j}$.
\begin{equation}
R_{i,j} = \Theta(\epsilon - ||\textbf{U}_i(d) - \textbf{U}_j(d)||),
\end{equation}
where, $\Theta$ is the Heaviside step function, $\epsilon$ is the recurrence threshold, and $|| ~. ~||$ is the Euclidean or maximum norm, and $\textbf{U}_i(d) = [u(i),u(i+\tau),...,u(i+\tau(d-1)]$ is the $i_{th}$ reconstructed phase-space vector. Here $\tau$ represents the optimal time delay and $d$ is the embedding dimension needed for a one-to-one projection of the original attractor. Any two states vector, $\textbf{U}_i$ and $\textbf{U}_j$ are considered to be neighbour if the distance between them in $d$-dimensional phase space is less than a threshold $\epsilon$: $||\textbf{U}_i(d)-\textbf{U}_j(d)||<\epsilon$. Thus, the element of the recurrence matrix $R_{i,j}=1$ if a state vector $\textbf{U}_i$ is a neighbour to state $\textbf{U}_j$ in phase space, otherwise $R_{i,j}=0$.



\subsection{\textbf{Recurrence quantification analysis}}
We obtain quantitative insight into nonlinear dynamics of the forced jet using recurrence quantification analysis (RQA)\cite{webber2005}. RQA computes statistical measures such as recurrence rate, determinism, and laminarity that are based on geometric patterns like diagonal or vertical lines in the RPs. These measures describe the geometrical characteristics of the underlying phase space attractor. We compute these RQA measure for smaller time windows in the recurrence matrix while moving along the main diagonal to derive time-varying RQA measures. In the present study, we use two RQA measures namely, recurrence rate and determinism to characterize the dynamical states in the lead up to lock-in.


\subsubsection{\textbf{Recurrence rate}}

Recurrence rate is defined as the density of recurring points in the recurrence plots.

\begin{equation}
RR = \frac{1}{N^2} \sum _{i,j=1}^{N} R_{i,j},
\end{equation}

Recurrence rate corresponds to the probability a state recurs to nearly same place i.e. within a distance of recurrence threshold $\epsilon$ in the phase space.

\subsubsection{\textbf{Determinism}}

Determinism ($DET$) measures the percentage of recurrence points in a recurrence matrix that form diagonal lines of a specified minimum length $l_{min}$ and is given as $DET$:
\begin{equation}
DET = \frac{\sum_{v=v_{min}}^{N}lP(l)}{\sum_{v=1}^{N}lP(l)},
\end{equation}

where, $P(ll)$ is the probability of existence of diagonal line of length $l$.

\subsection{\textbf{Joint recurrence analysis}}\label{sec4}

Joint recurrence plot is a bi-variate extension of a standard recurrence plots. JRPs highlight moments of simultaneous recurrence, allowing for the detection of synchronized behavior or shared dynamics. This method is useful for exploring relationships between disparate systems and can provide insights into complex interactions that are not apparent from analyzing each system individually.  

\begin{figure*}[h!]
	\centering
	\includegraphics[width=0.95\textwidth]{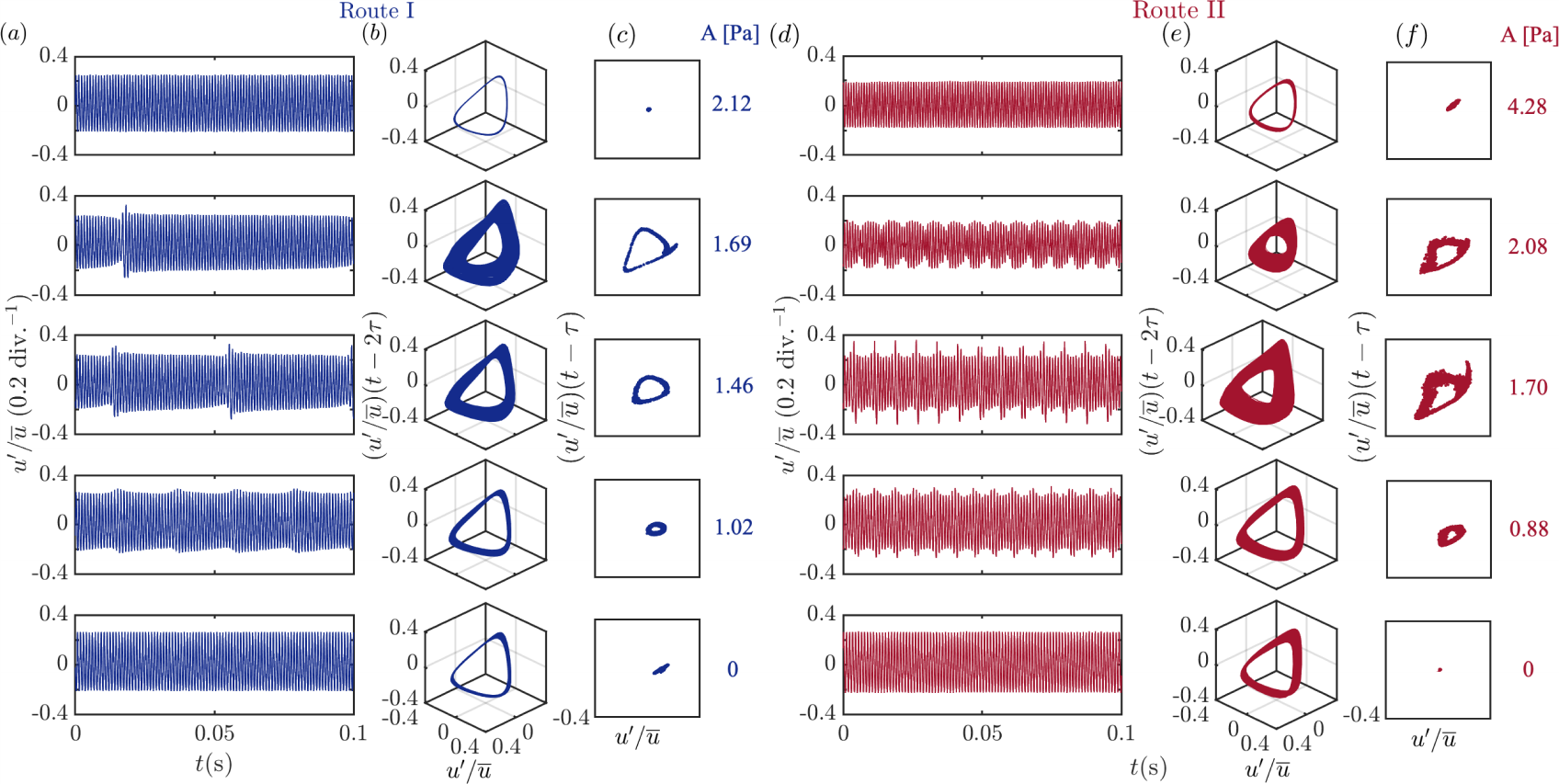}
	\caption{Time trace, phase space and Poincar\'{e} maps are shown for varying forcing amplitudes ($A$ [Pa]) as the low-density jet locks into the external forcing via ($a,b,c$) Route I (forcing close to the natural frequency, $f_f/f_n$ $\approx$ 1.04) and (d,e,f) Route II (forcing far away the natural frequency, $f_f/f_n$ $\approx$ 1.12) respectively.}
	\label{timetrace}
\end{figure*}

In joint recurrence plots, the individual phase spaces of the systems are preserved, and the comparison is made only at the time instants when both systems recur simultaneously, highlighting their joint recurrences. The joint recurrence of two systems, $\textbf{U}_i$ and $\textbf{V}_i$, can be mathematically represented in the form of,

\begin{equation}
JR_{i,j} = \Theta(\epsilon - ||\textbf{U}_i - \textbf{U}_j||)  \Theta(\epsilon - ||\textbf{V}_i - \textbf{V}_j||),
\end{equation}

The joint recurrence matrix, $JR_{i,j}$ takes a value of 1 when the phase space trajectory of one system $\textbf{U}_i$ recurs within a distance of a threshold $\epsilon$, while simultaneously the trajectory of the other system $\textbf{U}_j$ returns to its neighborhood in its own phase space. The recurrence thresholds may differ between systems, depending on the amplitude of data points in each time series.

Similar to standard RQA, statistical measures derived from $JR_{i,j}$ can help identify synchronization between interacting systems. In this study, we use joint recurrence rate ($jRR$) to identify the lock-in state of the forced jet. It can be directly estimated from the percentage of recurring points in the joint recurrence matrix.

\begin{equation}
jRR = \frac{1}{N^2} \sum _{i,j=1}^{N} JR_{i,j},
\end{equation}  

$jRR$ represents the joint correlation sum and it is expected to have higher values when two systems are synchronized.

\section{\textbf{RESULTS AND DISCUSSION}}\label{sec5}


Figure ~\ref{timetrace} shows synchronization dynamics and a sequence of bifurcations leading to lock-in when a self-excited jet is axially forced with progressively increasing forcing amplitudes. We consider two forcing frequencies: (i) route I: forcing close to the natural frequency ($f_f/f_n$ $\approx$ 1.04) and (ii) route II: forcing away from the natural frequency ($f_f/f_n\approx 1.12$). For both forced cases, we examine the time traces, the phase space and the one-sided Poincar\'{e} map of normalized unsteady velocity ($u'/\overline{u}$) acquired from the hydrodynamically self-excited, low-density jet. 
The unforced jet ($A=0$ [Pa] for both $f_f/f_n$ $\approx$ 1.04 and 1.12 in Fig.\ref{timetrace}) exhibits self-excited limit cycle oscillations at a discrete natural frequency ($f_n=945$ Hz). The limit cycle oscillations is evidenced in their respective phase spaces (Fig.\ref{timetrace} b,e) where the phase trajectory evolves around a closed repetitive orbit. Both Poincar\'{e} maps (Fig.\ref{timetrace} c,f) show intercepts of the phase trajectory concentrated as single cluster of point and the absence of amplitude modulation in the time traces (Fig. \ref{timetrace} a,d) further confirms that the unforced jet is at period-1 state. When  forced at moderate amplitudes ($A=1.02-1.69$ [Pa] for $f_f/f_n$ $\approx$ 1.04; $A=0.88-2.08$ [Pa] for $f_f/f_n \approx 1.12$), the jet transitions to a quasiperiodic state. This is indicated by the presence of amplitude modulation in the time trace and the emergence of a closed continuous ring of trajectory intercepts in the Poincar\'{e} map. Collectively, these observations suggest that the jet shifts from a period-1 limit cycle attractor to a two-dimensional $\mathbb{T}^2$ torus attractor via a Neimark-Sacker bifurcation. With further increase in A, a reduction in the beating frequency of amplitude modulation is seen for $f_f/f_n$ $\approx$ 1.04, while the beating frequency remains constant for $f_f/f_n$ $\approx$ 1.12. A reduction in the beating frequency for $f_f/f_n\approx 1.04$ indicates a pulling phenomenon, which is a hallmark of phase-locking route to synchronization. Furthermore, the phase trajectory continues to evolves around the ergodic $\mathbb{T}^2$ torus attactor while the torus grows monotonically with A for $f_f/f_n\approx 1.04$. By contrast, for $f_f/f_n\approx 1.12$, the size of the torus attractor is seen to first grow, reaches peak at $A$=1.70, then gradually shrink with increment in $A$ as evidenced in the phase space and in the Poincar\'{e} map. When forced at a critical amplitude ($A=2.12$ [Pa] for $f_f/f_n$ $\approx$ 1.04; and $A=4.28$ [Pa] for $f_f/f_n$ $\approx$1.12 in 

\begin{figure*}[h!]
	\centering	\includegraphics[width=0.99\textwidth]{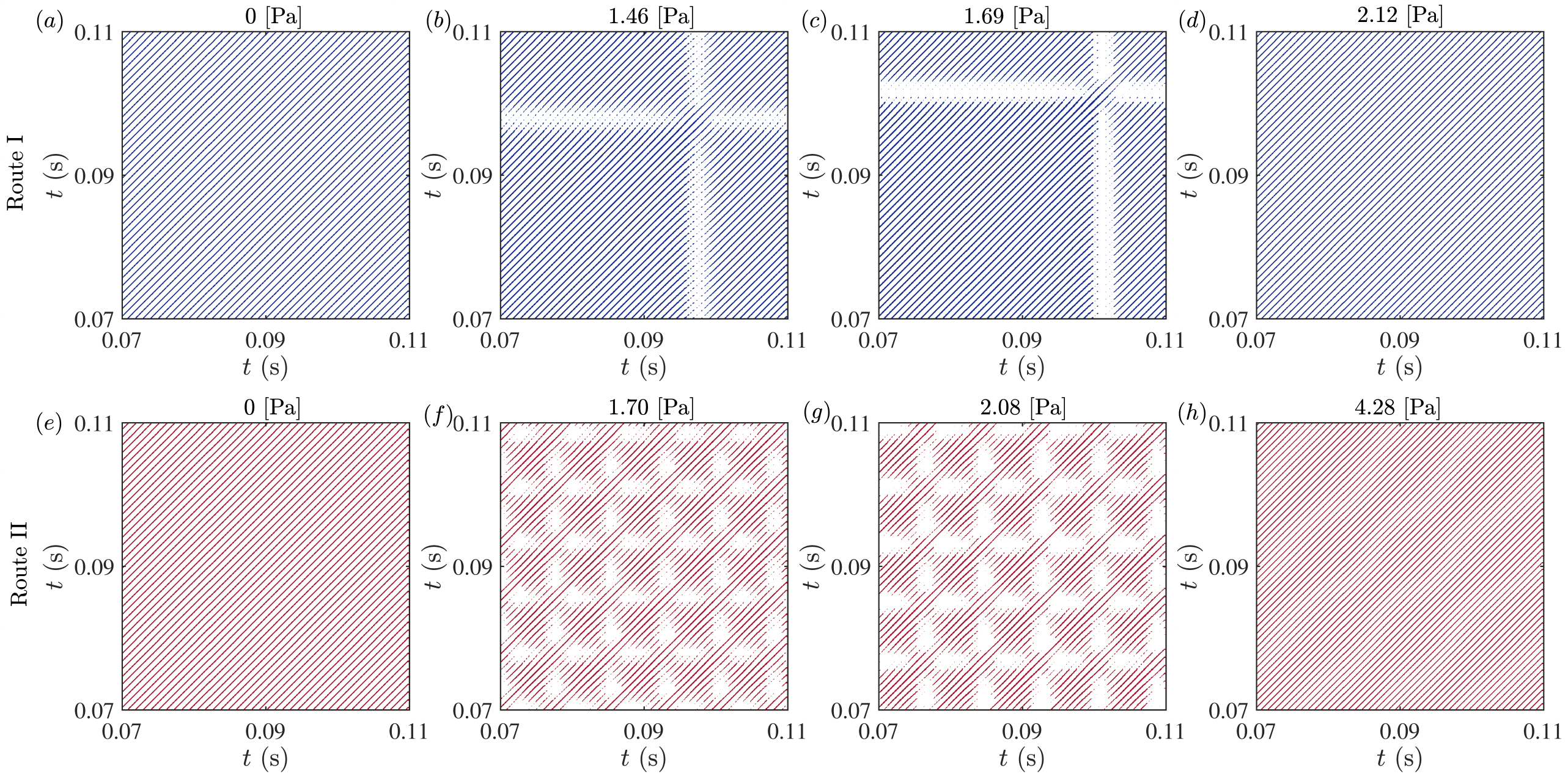}
	\caption{Recurrence plots for the self-excited limit cycle, quasiperiodicity and lock-in states are shown whem ($a-d$) the frequency is close to the natural frequency ($f_f/f_n$ $\approx$ 1.04) and ($e-h$) far away the natural frequency ($f_f/f_n$ $\approx$ 1.12) respectively. }
	\label{RPshortseries}
\end{figure*}

Fig.\ref{timetrace}), the jet lock-in to $f_f$. Consequently, the amplitude modulation disappears in the time trace for both forced cases. During the transition from quasiperiodicity to lock-in, the ergodic $\mathbb{T}^2$ attractor undergoes a distinct dynamical changes depending on whether $f_f$ is close to $f_n$ or far from it. Specifically, the torus attractor abruptly collapse into a stable periodic orbit when $f_f$ is close to $f_n$ ($f_f/f_n$ $\approx$ 1.04) as seen in phase space. This abrupt collapse is unique feature of saddle-node bifurcation and it is evidenced by sudden collapse of ring structures into a single clusters trajectory intercepts in the Poincar\'{e} map. When $f_f$ is far from $f_n$ ($f_f/f_n\approx1.12$), the torus attractor collapses gradually leading to its eventual disappearance through an inverse Neimark-Sacker (torus-death) bifurcation, as seen in the phase space and in the Poincar\'{e} map. This route to lock-in also results in a small reduction in the overall oscillations amplitude at lock-in compared to the unforced case.        


Next, we transform the time series data into two-dimensional binary bitmap data using recurrence plots, with the goal of applying recurrence quantification analysis (RQA) to identify quantitative measures that can distinguish and predict the two routes to synchronization.

\subsection{\textbf{Recurrence analysis }}

Figure \ref{RPshortseries} shows recurrence plots for $u^\prime/\overline{u}$ signals for the jet exhibiting (i) periodic oscillations when unforced (Fig. \ref{RPshortseries}a,e), (ii) quasiperiodic oscillations when forced at moderate amplitude (Fig. \ref{RPshortseries}b,c,f,g), and (iii) periodic oscillations at lock-in (Fig. \ref{RPshortseries}d,h) for two forcing conditions: $f_f/f_n \approx 1.04$ (Route I) and $f_f/f_n\approx 1.12$ (Route II). We choose the recurrence thresholds ($\epsilon$) as one fourth of the size of the corresponding phase space attractors.

\begin{figure}
	\centering
	\includegraphics[width=0.99\textwidth]{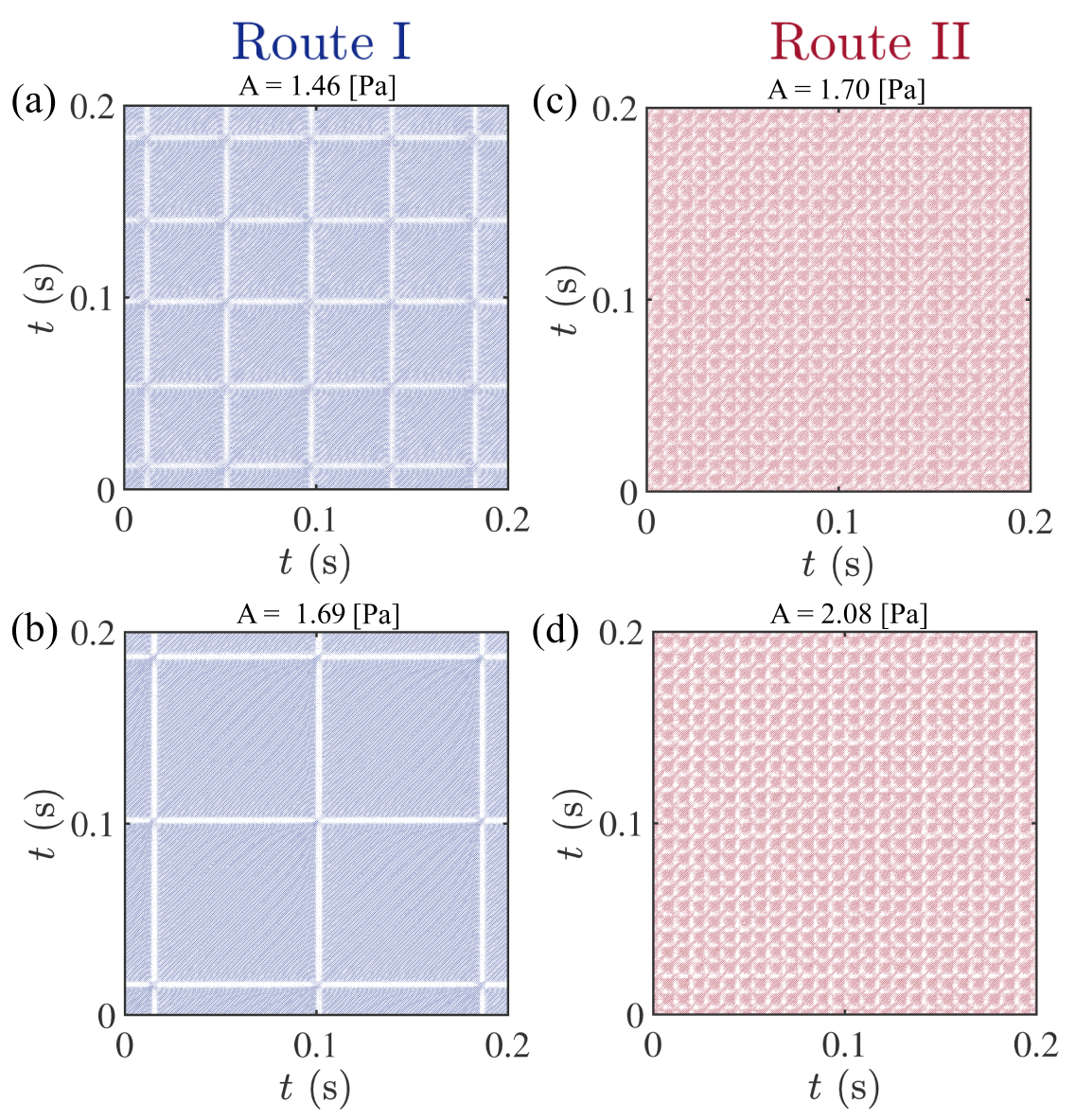}
	\caption{The same as Fig. \ref{RPshortseries}(b,c,f,g) but for a longer time duration.}
	\label{RPlongseries}
\end{figure}

The recurrence plot for the jet exhibiting period-1 motion is seen as pattern of equally-spaced continuous diagonal lines. This is evidenced by Fig. \ref{RPshortseries}(a,d,e,h) where long non-interrupted diagonal lines are visible regardless of whether the periodic state arises due to lock-in or in unforced condition. The vertical distance between the diagonal lines in the recurrence plot signifies the time period of oscillations. Consequently, aoong route II, the two consecutive diagonal lines in the lock-in state are closer together than in the unforced case. During quasiperiodic state, low-frequency amplitude modulations cause interruptions in the long diagonal lines, leading to the appearance of non-recurring white spaces in the recurrence plots. This is visualized by Fig. \ref{RPshortseries}(b,c,f,g). Notably, the pattern of recurring diagonal lines depends on the type of route taken to lock-in. The recurrence plot of the quasiperiodic attractor along route I have higher density of longer diagonal lines. By contrast, a higher density of disconnected and short segments of diagonal lines  are visible in the recurrence plots along route II. Physically, this implies that the jet exhibiting quasiperiodic oscillations along route I is more periodic for longer epoch, indicating a low frequency amplitude modulation. Along route II, however, the jet is highly aperiodic with strong amplitude modulation. To further corroborate that the quasiperiodically oscillating jet along route I indeed spend more time oscillating periodically as it approach lock-in, we show in Fig. \ref{RPlongseries} the recurrence plots with a higher number of data points in the respective recurrence matrices, which correspond to the time duration of approximately 0.2 seconds. 

This allows us to visualize the dynamics with both the fast (of the order of natural frequency of self-excited limit cycle) and slow time scales (of the order of low-frequency modulations in the quasiperiodic attractor). As discussed before, a higher density of relatively long discontinuous diagonal lines in the recurrence plots are evident for the quasiperiodically oscillating jet aong route I (Fig. \ref{RPlongseries}a,b). The length of these broken diagonal segments is seen to increases with increase in the forcing amplitude, suggesting presence of longer epoch of periodic oscillations. Furthermore, the distance between the white lines of finite-width which marks the time scale of the amplitude modulations in the quasiperiodic time series also seen to increase as forcing amplitude approach towards the critical lock-in amplitude (Fig. \ref{RPlongseries}a,b). In a stark contrast to route I, the recurrence plot (Fig. \ref{RPlongseries}c,d) show no change in the density of broken, short diagonal lines with increases in the forcing amplitude along route II. The presence of high density of short diagonal lines with more frequent interruption of non-recurrence white patterns constitute a corroborative evidence of highly aperiodic oscillations. Notably, the equispaced broken short diagonal lines and finite-width white spaces in the recurrence plots (Fig. \ref{RPlongseries}c,d) further indicates that the underlying dynamics exhibits a constant beating frequency of amplitude modulation. Collectively, these observations imply that along route II, the jet continues to oscillates at a constant beat frequency as it approach lock-in amplitude without changes in the beating frequency.


\begin{figure}
	\centering
	\includegraphics[width=0.99\textwidth]{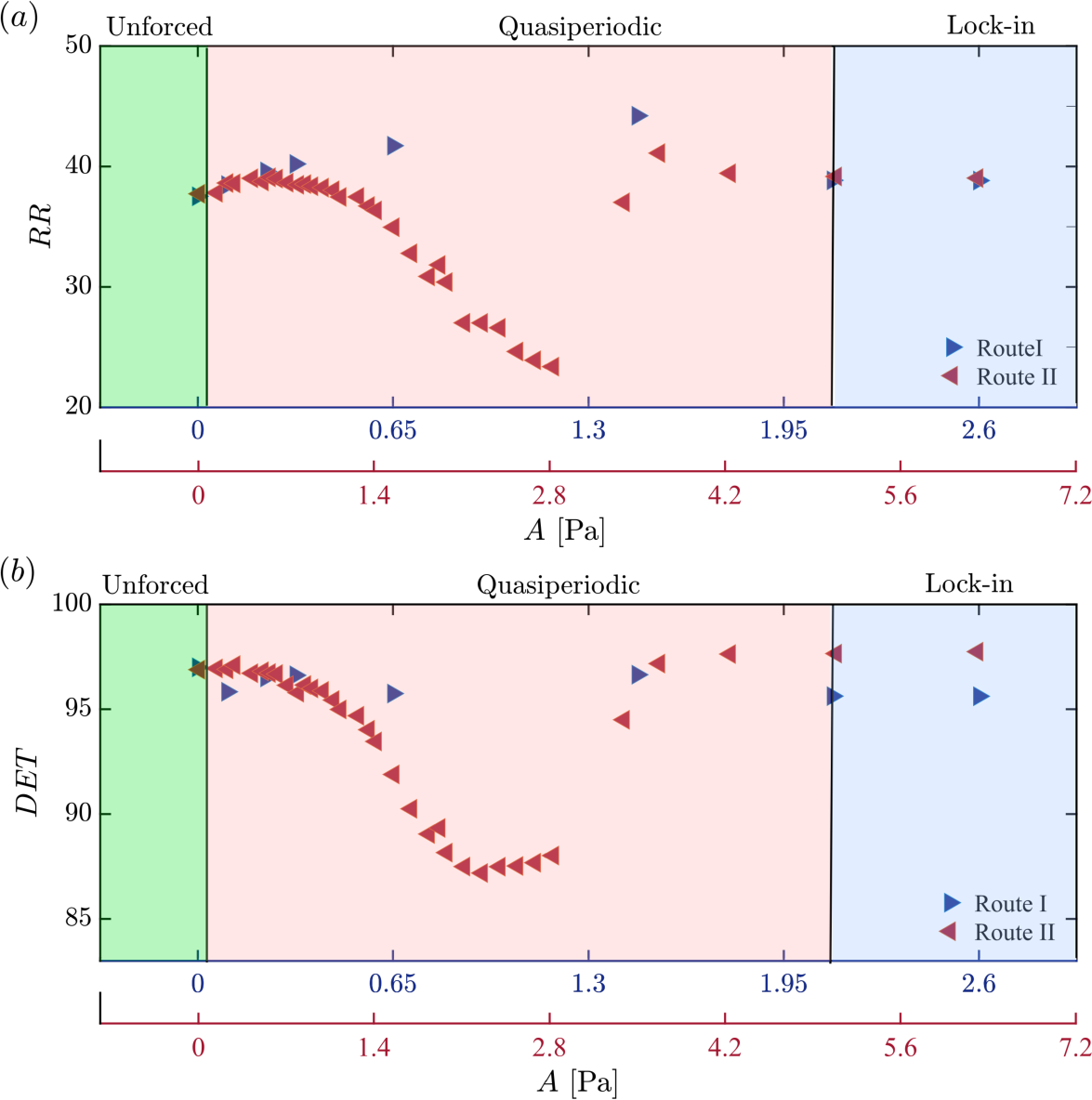}
	\caption{Variation of RR (a) and DET (b) as functions of A for both saddle-node (route I) and torus-death bifurcation route (route II).}
	\label{RQA}
\end{figure}

\subsection{\textbf{Detecting Synchronization using RQA}}

Figure \ref{RQA} shows variation of $RR$ and $DET$ computed from recurrence plot as functions of forcing amplitude A for both routes to synchronization: route I and route II. We find that both $RR$ and $DET$ exhibit similar qualitative behavior as $A$ increases. Along route I, the values of $RR$ and $DET$ remain relatively constant. However, along route II, both measures vary non-monotonically with increasing $A$. Specifically, they initially decrease under weak forcing, reach minimum values at moderate forcing amplitudes, and then gradually rise, approaching the values of the unforced state as the forcing amplitude nears the critical lock-in threshold. A relatively higher values of $RR$ and $DET$ along route I compared to route II in the quasiperiodic region implies that the time trace of jet oscillations is more periodic and the dynamics are more deterministic and stable along the saddle-node bifurcation route compared to the torus-death bifurcation route.

\begin{figure}
	\centering
	\includegraphics[width=0.99\textwidth]{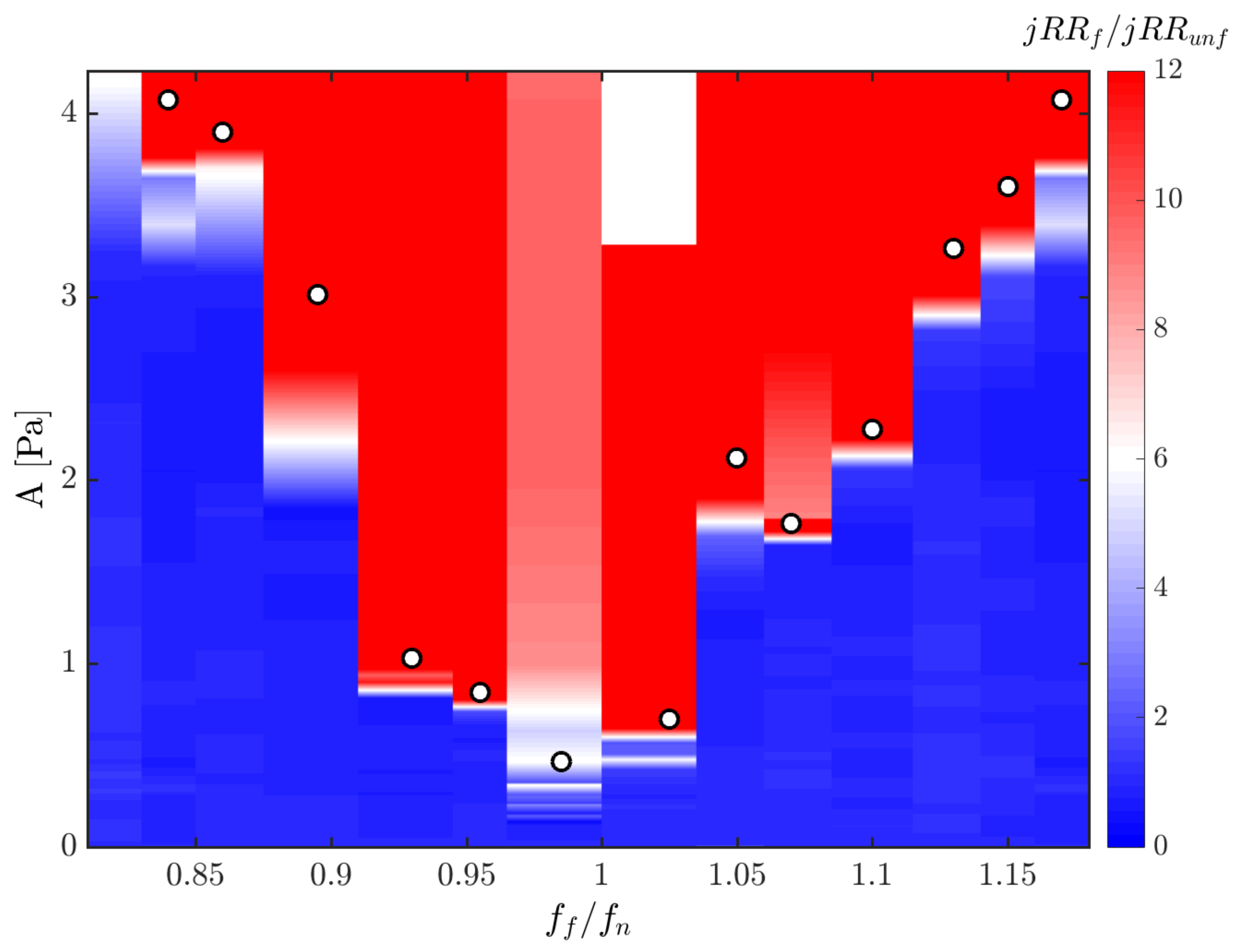}
	\caption{Synchronization map of axially forced self-excited jet showing contours of the normalized joint recurrence rate ($jRR$) in parameter space defined by the forcing amplitude ($A$) and the normalized forcing frequency ($f_f/f_n$. The discrete circular markers indicate onset of synchronization.}
	\label{jRR}
\end{figure}

Figure \ref{jRR} shows the synchronization map of axially forced jet in a parameter space defined by the
forcing frequency ($f_f/f_n$) and the forcing amplitude ($A)$. This maps displays contours of normalized joint recurrence rate ($jRR_f/jRR_{unf}$) computed from joint recurrence plot. The discrete circular makers overlaid on the contours plots indicates the onset of synchronization. 
The map reveals that the joint recurrence rate ($jRR$) increases with $A$, reaching its maximum when the jet locks in to $f_f$, regardless of whether $f_f$ is close to or far from $f_n$. This is because at lock-in, the simultaneous recurrences in the JRP become more frequent, resulting in a rise in $jRR$. 

Crucially, the statistical measures derived from the standard recurrence plots, combined with those from joint recurrence plots, are highly effective in distinguishing the unique bifurcation route in lead up to lock-in.

\section{\textbf{CONCLUSIONS}}\label{sec4}
In the paper, we have investigated the synchronization dynamics and analyzed the distinct routes to synchronization in a hydrodynamically self-excited low-density jet. By leveraging statistical measures derived from both standard and joint recurrence plots, we have successfully differentiated between the saddle-node and torus-death routes to synchronization. The recurrence rate ($RR$) and determinism ($DET$) computed from the recurrence plots, combined with with joint recurrence rate ($jRR$) from joint recurrence plots, have proven to be effective tools for identifying and forecasting the proximity of a self-excited system to its lock-in boundaries. These findings not only advance our understanding of forced synchronization in self-excited systems but also offer valuable insights for developing low-order models to predict and control such systems.

\vspace{0.5cm}
\noindent
\textbf{ACKNOWLEDGEMENTS}\\
\noindent Author (Meenatchidevi Murugesan) acknowledges financial support from the Science and Engineering Research Board (SERB) of India, granted under No. CRG/2023/001277 and CRG/2023/004134.     \\
\vspace{0.5cm}
 \noindent

\bibliographystyle{unsrtdin}
\bibliography{FMFP2024}

\end{document}